\documentclass[twocolumn,secnumarabic,superscriptaddress,amssymb, nobibnotes, aps, prl]{revtex4-2} 
\usepackage{epsfig,graphicx,subfigure,hyperref}
\hypersetup{
    colorlinks=true,
    linkcolor=blue,
    filecolor=blue,      
    urlcolor=blue,
    citecolor=blue
    }

\usepackage{amsmath,ulem,mathrsfs}
\usepackage{xcolor}

\newcommand{\V}{\boldsymbol}

\begin{document}

\title{Multipolar Light–Matter Hamiltonians in Symmetry-Breaking Photonic Vacuums}

\author{Liu Yang}\thanks{These two authors contributed equally}
\affiliation{Tsung-Dao Lee Institute, Shanghai Jiao Tong University, Shanghai, 201210, China}
\affiliation{School of Physics and Astronomy, Shanghai Jiao Tong University, Shanghai 200240, China}
\author{Jiadu Lin}\thanks{These two authors contributed equally}
\affiliation{Tsung-Dao Lee Institute, Shanghai Jiao Tong University, Shanghai, 201210, China}
\affiliation{School of Physics and Astronomy, Shanghai Jiao Tong University, Shanghai 200240, China}
\author{Qing-Dong Jiang}
\email{qingdong.jiang@sjtu.edu.cn}
\affiliation{Tsung-Dao Lee Institute, Shanghai Jiao Tong University, Shanghai, 201210, China}
\affiliation{School of Physics and Astronomy, Shanghai Jiao Tong University, Shanghai 200240, China}
\affiliation{Shanghai Branch, Hefei National Laboratory, Shanghai 201315, China}

\begin{abstract}

We show that the conventional multipolar Hamiltonian is qualitatively modified when the photonic vacuum breaks inversion or time-reversal symmetry. By explicitly applying the Power-Zienau-Woolley transformation, we derive the resulting multipolar Hamiltonians for two idealized chiral photonic environments: a spatial-chiral vacuum, which breaks inversion symmetry, and a temporal-chiral vacuum, which breaks time-reversal symmetry. In the spatial-chiral case, the transformation generates an inversion-breaking self-energy, whereas in the temporal-chiral case it produces an {additional Zeeman-like energy}. Using a trapped hydrogen-like atom and a charged harmonic oscillator in cavities as minimal examples, we show that these symmetry-dependent terms lead to characteristic spectral shifts. Our work provides a general framework for describing light-matter interactions in chiral quantum electrodynamics and identifying the associated symmetry-dependent effects on cavity-embedded atoms, molecules, and quantum materials.

\end{abstract}
\maketitle

\textit{Introduction}.—Vacuum fluctuations of the electromagnetic field provide a powerful route for controlling atoms, molecules, and quantum materials~\cite{Molecular2016,RMP2019,NatureReview2019,MarkFox2023,Molecular2025,RevQEDgas2021,Manipulating2021,review_cavity2022,cavityQHE2022,cavityHall2025nature,Perspective2025}. Embedding matter in a small cavity enhances light-matter coupling and thereby enables access to the strong-coupling regime, qualitatively reshaping collective excitations and stabilizing new quantum phases~\cite{cohen3,Haroche2006,Aspect2010,RMP2019,NatureReview2019,Ma2023,Ma2026,Fu2026}. More broadly, cavity design provides an increasingly versatile means of engineering not only the strength, but also the mode structure and symmetry, of the photonic vacuum. Recent advances in chiral mirrors, nanophotonic resonators, microcavities, and terahertz photonic-crystal cavities have enabled photonic environments with controlled helicity and broken spatial or temporal symmetries~\cite{chiralmirrors2015,chiralnano2019,Baranov2020,chiral_cavity2021,chiralQED2022,NCchiralcavity2024,Terahertzchiral2024,Terahertzchiral2025,2026yaowang,bchn-b47c,chiral_TRS2026}. These developments have stimulated growing interest in symmetry-breaking cavity quantum electrodynamic (QED) and its effects on embedded matter. In molecular systems, chiral cavities can induce enantioselective interactions and chirality-dependent spectral shifts~\cite{chirality_select2024,Nonperturbative2023,jiang2023}. In quantum materials, they offer new possibilities for engineering electronic topology and electromagnetic responses, including in graphene and quantum Hall systems~\cite{yang2024emergent,jiang_angular2025,Liuchiral2024}.

A fundamental question is therefore how light-matter interactions should be formulated in such symmetry-breaking photonic vacuums. The canonical description starts from the nonrelativistic minimal-coupling Hamiltonian~\cite{cohen3}. Although formally exact, this representation expresses the interaction through the gauge-dependent vector potential and can obscure its physical multipolar structure. The Power-Zienau-Woolley (PZW) transformation instead recasts the Hamiltonian in terms of electric and magnetic multipoles coupled directly to the physical electromagnetic fields, providing a transparent description of localized atoms and molecules~\cite{PZ1957,power1959coulomb,atkins1970interaction,Babiker1974O,Power1982,BabikerLoudon1983,Stokes_2018,PZW2020,StokesRMP,Stokes2023,cohen3,cohen4,Babiker2024Gauge}. In conventional photonic vacuums preserving both inversion ($\mathcal{I}$) and time-reversal ($\mathcal{T}$) symmetries, this transformation yields the familiar multipolar Hamiltonian. Whether this conventional form remains complete when the photonic vacuum itself breaks $\mathcal{I}$ or $\mathcal{T}$ symmetry, however, has not been established.

To answer this question, we first derive the multipolar light-matter Hamiltonian for a general photonic vacuum without assuming inversion or time-reversal symmetry. We then specialize this general result to two idealized chiral photonic vacuums: a spatial-chiral vacuum, in which counterpropagating modes possess the same helicity, and a temporal-chiral vacuum, in which counterpropagating modes share the same laboratory-frame circular polarization.  We find that the resulting multipolar Hamiltonians acquire new symmetry-breaking terms beyond the conventional PZW Hamiltonian. In the spatial-chiral case, the PZW transformation generates an inversion-breaking self-energy, whereas in the temporal-chiral case it produces a cavity-induced Zeeman-like term. Finally, we illustrate the spectral consequences of these additional terms using two representative models: a trapped hydrogen-like atomic system in a spatial-chiral cavity and a charged two-dimensional harmonic oscillator coupled to a temporal-chiral cavity mode. 
Together, these results reveal how broken inversion or time-reversal symmetry of the photonic vacuum is encoded in the multipolar Hamiltonian and gives rise to distinct spectral signatures in embedded matter.

\textit{PZW Hamiltonian in a general photonic vacuum}.—The canonical Hamiltonian of non-relativistic QED that describes the interaction between charged particles and the quantized electromagnetic field in the Coulomb gauge is given by~\cite{cohen3}
\begin{align}
    H&=\sum_\alpha\frac{[\boldsymbol{p}_\alpha-q_\alpha\boldsymbol{\hat{A}}(\boldsymbol{r}_\alpha)]^2}{2m_\alpha}+ H_\text{Cou}+ H_\text{rad},\label{eq:total_H}
\end{align}
where $\boldsymbol{p}_\alpha$ is the canonical momentum operator of particle $\alpha$, $H_\text{Cou}$ represents the Coulomb interaction, $\boldsymbol{\hat{A}}$ is the vector potential and $ H_\text{rad}$ is the radiation field energy. The Hamiltonian of the radiation field is quantized as $ H_\text{rad}=\sum_{\ell}\hbar\omega_{\ell}(a_\ell^\dagger a_\ell+\frac{1}{2})$ with $[a_\ell,a^\dagger_\ell]=1$, and the vector potential is defined as
\begin{align}
\boldsymbol{\hat{A}}(\boldsymbol{x})&=\sum_{\ell}\mathcal{A}_\ell [a_\ell \V{f}_\ell(\boldsymbol{x} )+a_\ell^\dagger \V{f}^\ast_\ell(\boldsymbol{x} )],
\end{align}
where  $\omega_\ell$ is the frequency of the mode $\ell$ and the field amplitude is given by $\mathcal{A}_\ell =\sqrt{\hbar/(2\varepsilon_0 V_\ell\omega_\ell)}$ ($V_\ell$ is the associated mode volume). In this work, we consider general photonic vacuums in which inversion or time-reversal symmetry may be broken. For convenience, we introduce the following general commutation relations between different components of vector potential and magnetic field:
\begin{align}
    [\hat{A}^i(\V{x}),\hat{A}^j(\V{x}')]&=i\hbar\Delta^{ij}(\V{x},\V{x}'),\\
    [\hat{A}^i(\V{x}),\hat{B}^j(\V{x}')]&=i\hbar\Gamma^{ij}(\V{x},\V{x}').
\end{align}

The PZW transformation provides a way to re-express the interaction in terms of multipole-field couplings in the Hamiltonian~\cite{PZ1957,power1959coulomb,atkins1970interaction,Babiker1974O,Power1982,BabikerLoudon1983,Stokes_2018,PZW2020,StokesRMP,Stokes2023,cohen4,Babiker2024Gauge}. The PZW transformation
$H_{\rm P}=UHU^\dagger$ is generated by the unitary operator $U=\exp\big(-\frac{ i }{\hbar}\int d^3 x\boldsymbol{P}(\boldsymbol{x})\cdot\boldsymbol{\hat{A}}(\boldsymbol{x})\big)$, where $\boldsymbol{P}(\boldsymbol{x})$ is the polarization density. One convenient definition introduces a continuous distribution of dipoles along the straight line connecting the reference point to the particle's position $\boldsymbol r_\alpha$~\cite{cohen4,PZW2020}: 
\begin{align}
    \boldsymbol{P}(\boldsymbol{x})=\int_0^1 du \sum_\alpha q_\alpha\boldsymbol{r}_\alpha\delta(\boldsymbol{x}-u\boldsymbol{r}_\alpha).\label{eq:P_u}
\end{align}
We emphasize that the PZW transformation is defined with respect to a chosen reference point. Therefore, translational symmetry is no longer manifest in the multipolar representation even for plane-wave electromagnetic modes. After the transformation, we obtain the PZW Hamiltonian as follows
\begin{align}
H_{\rm P}=&\sum_\alpha\frac{[ \V{p}_\alpha-q_\alpha  \V{\hat{A}}_{\rm P}(\V{r}_\alpha)-q_\alpha \delta \V{A}_{\text{P} }(\{\V{r}_\alpha\})]^2}{2m_\alpha}\nonumber\\&+ \mathcal{E}^\text{Cou}+H^\text{rad}-\int d^3x \,\V{P}(\V{x})\cdot\V{\hat{E}}(\V{x})\nonumber\\&+\sum_{\ell}\frac{\mathcal{A}_\ell^2\omega_\ell}{\hbar}|\int_0^1 du \sum_\alpha q_\alpha \V{r}_\alpha\cdot\V{f}_\ell(u\V{r}_\alpha)|^2,\label{eq:generalPZW}
\end{align}
where the quantized vector potential
\begin{align}
   \boldsymbol{\hat{A}}_{\rm P}(\boldsymbol{x})&= \int_0^1 udu \,\hat{\boldsymbol{B}}(u\boldsymbol{x})\times \boldsymbol{x},
\end{align}
satisfies the so-called Poincaré gauge $\boldsymbol{x} \cdot   \boldsymbol{\hat{A}}_{\rm P}(\boldsymbol{x}) = 0$ accompanied with a classical vector potential 
\begin{align}
  &\delta A_{\text{P} }^i(\{\V{r}_\alpha\}) =\frac{1}{2}\int_0^1 du\sum_\beta q_\beta r^j_\beta \Delta^{ji}(u\V{r}_\beta,\V{r}_\alpha)\nonumber\\
    &+\frac{1}{2}\int_0^1 dv\int_0^1 u du\sum_\beta q_\beta r^j_\beta r^l_\alpha \epsilon^{i k l }\Gamma^{jk}(v \V{r}_\beta,u \V{r}_\alpha).
\end{align}
Here, $\delta\boldsymbol{A}_{\rm P}(\{\boldsymbol{r}_\alpha\})$ denotes the vector potential acting on particle $\alpha$, while its value depends on the full particle configuration. The derivation of the PZW transformation, including its extension to spinful particles, is provided in the Supplemental Material~\cite{supplementary}. Eq.~(\ref{eq:generalPZW}) shows that, while the Poincaré-gauge vector potential $\boldsymbol{\hat{A}}_{\rm P}$, multipole-field coupling, and polarization self-energy (the last term in Eq.~(\ref{eq:generalPZW})) retain the conventional PZW structure~\cite{PZW2020,StokesRMP,supplementary}, broken time-reversal symmetry can additionally generate the vector potential $\delta\boldsymbol{A}_{\rm P}$ in the kinetic momentum. Moreover, because the explicit form of the self-energy depends on the photonic mode structure, it can contain symmetry-breaking contributions. 

\begin{figure}
    \centering
    \includegraphics[width=0.6\linewidth]{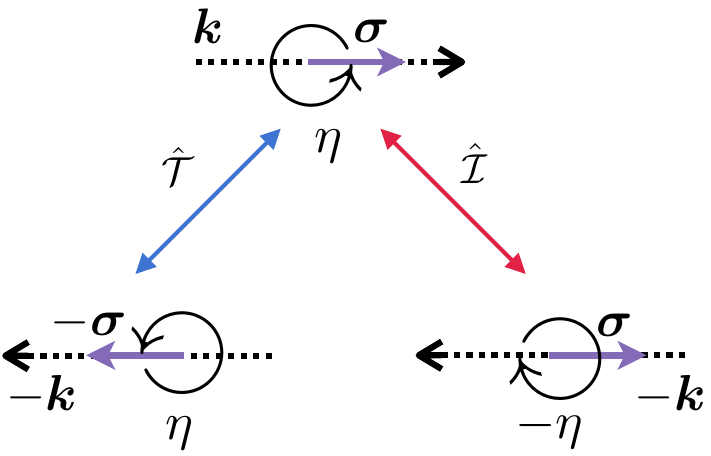}
    \caption{Illustration of inversion ($\hat{\mathcal{I}}$) and time-reversal ($\hat{\mathcal{T}}$) transformations for a plane-wave mode with wavevector $\boldsymbol{k}$, (dimensionless) spin angular momentum $\V{\sigma}$, and helicity $\eta=\boldsymbol{k}\cdot\V{\sigma}/|\boldsymbol{k}|$.}
    \label{fig:transform}
\end{figure}

\textit{Ideal spatial- and temporal-chiral vacuums}.—In this work, we propose two special photonic vacuums that either break inversion or time-reversal symmetry. It is useful to first examine how individual photon modes transform under these symmetries. Under spatial inversion $\hat{\mathcal{I}}$, the wavevector $\boldsymbol{k}$ is inverted, and the helicity $\eta$ also flips sign. Under time reversal $\hat{\mathcal{T}}$, the wavevector $\boldsymbol{k}$ is inverted but the helicity $\eta$ remains unchanged. Fig.~\ref{fig:transform} summarizes the corresponding transformations.

\begin{figure}
    \centering
    \includegraphics[width=0.98\linewidth]{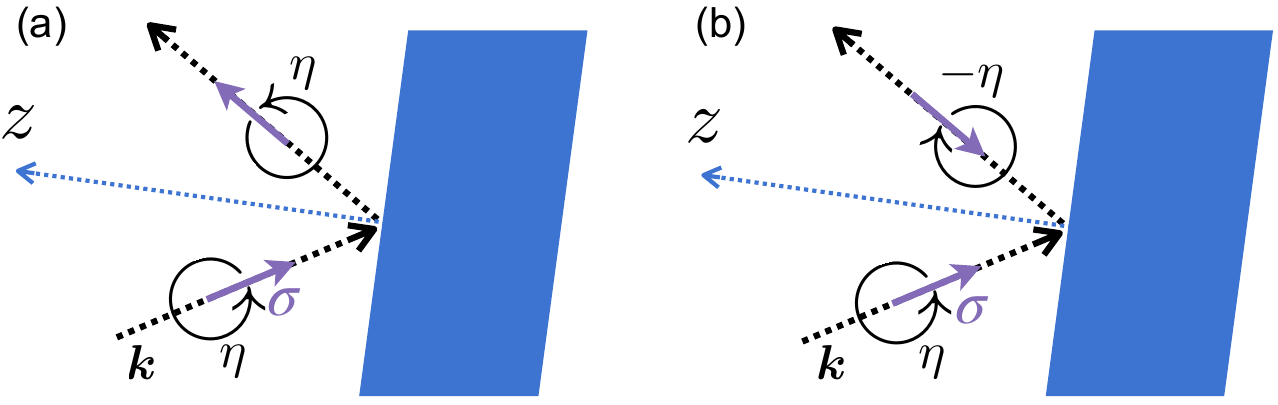}
    \caption{Schematic mirror configurations for two symmetry-breaking photonic vacuums.
(a) Spatial-chiral vacuum: both mirrors reflect the same helicity sector of modes
while preserving helicity and are transparent to the opposite sector.
(b) Temporal-chiral vacuum: the two mirrors reflect opposite helicity
sectors, reverse the helicity upon reflection, and are each transparent
to the other sector.}
    \label{fig:mirrors}
\end{figure}

First, we consider a spatial-chiral vacuum in which the $\boldsymbol{k}$ and $-\boldsymbol{k}$ modes have the same helicity $\eta$, thereby preserving $\mathcal{T}$ symmetry while breaking $\mathcal{I}$ symmetry. Its vector potential is
\begin{align}
    \boldsymbol{\hat{A}}^\eta(\boldsymbol{x})&=\sum_{\boldsymbol{k}}\mathcal{A}_k [a_\eta(\boldsymbol{k})\boldsymbol{\epsilon}_\eta(\boldsymbol{k} )e^{i\boldsymbol{k}\cdot\boldsymbol{x}}+\text{h.c}],\label{eq:A_I}
\end{align}
where the mode amplitude is given by $\mathcal{A}_k =\sqrt{\hbar/(2\varepsilon_0 V\omega_k)}$ with $\omega_k=|\boldsymbol{k}|c$, and $\boldsymbol{\epsilon}_\eta(\boldsymbol{k} )=(\boldsymbol{k}_1-i\eta\boldsymbol{k}_2)/\sqrt{2} $ is the polarization vector. Here, $\{\boldsymbol{k}_1,\boldsymbol{k}_2,\boldsymbol{k}_3=\boldsymbol{k}/|\boldsymbol{k}|\}$ forms a right-handed orthonormal basis. This spatial-chiral vacuum has been proposed to describe chiral cavities in Ref.~\cite{Nonperturbative2023}. It can be realized using helicity-selective mirrors, as illustrated in Fig.~\ref{fig:mirrors}(a). The corresponding PZW Hamiltonian is
\begin{align}
H_{\rm P}^\eta&=
     \sum_\alpha\frac{[ \V{p}_\alpha-q_\alpha  \V{\hat{A}}^\eta_{\rm P}(\V{r}_\alpha)]^2}{2m_\alpha}+ \mathcal{E}^\text{Cou}+H^\text{rad}\nonumber\\&-\int d^3x  \V{P}_\perp(\V{x})\cdot\V{\hat{E}}^\eta(\V{x})+\sum_{\boldsymbol{k}}\frac{\mathcal{A}_k ^2\omega_k}{2\hbar}|\widetilde{\V{P}}_\perp(\boldsymbol{k})|^2\nonumber\\&+i\eta\sum_{\boldsymbol{k}}\frac{\mathcal{A}_k ^2\omega_k}{2\hbar|\boldsymbol{k}|}\boldsymbol{k}\cdot[\widetilde{\V{P}}^\ast(\boldsymbol{k})\times\widetilde{\V{P}}(\boldsymbol{k})],\label{eq:spatialPZW}
\end{align}
where $\widetilde{\boldsymbol{P}}(\boldsymbol{k})=\int d^3x\,\boldsymbol{P}(\boldsymbol{x}) e^{-i\boldsymbol{k}\cdot\boldsymbol{x}}$ , and the subscript $\perp$ denotes the transverse component. Compared with the conventional PZW Hamiltonian~\cite{PZW2020,StokesRMP,supplementary}, Eq.~(\ref{eq:spatialPZW}) contains an additional inversion-breaking self-energy given by the last term. Expanding this self-energy to third order in the particle coordinates gives
\begin{align}
    \Delta H^\eta
    \simeq
   -\boldsymbol{d}\cdot\sum_{\boldsymbol{k}}\frac{\eta \mathcal{A}_k ^2\omega_k}{2\hbar|\boldsymbol{k}|}(\boldsymbol{k}\times\boldsymbol{Q}\cdot\boldsymbol{k}),
    \label{eq:corr_I}
\end{align}
where $\boldsymbol{Q}=\sum_\alpha q_\alpha\boldsymbol{r}_\alpha\boldsymbol{r}_\alpha$ is the total quadrupole-moment tensor. For particles with the same charge
$q$ and mass $m$, we define the mode-resolved coupling strength $g_k=|q|\mathcal{A}_k\sqrt{\omega_k/(\hbar m)}$~\cite{RMP2019,Ashida2021}. The contribution of each mode to $\Delta H^\eta$ therefore scales as $g_k^2\propto V^{-1}$.

Next, we consider a temporal-chiral vacuum in which the counterpropagating modes share the same
laboratory-frame circular polarization. This vacuum breaks $\mathcal{T}$ symmetry while preserving $\mathcal{I}$ symmetry~\cite{chiral_cavity2021,Terahertzchiral2025,chiral_TRS2026}. Such a vacuum can be described by the following vector potential:
\begin{align}
\boldsymbol{\hat{A}}^{\zeta}(\boldsymbol{x})=\sum_{\boldsymbol{k}}&\mathcal{A}_k [ a_{ \text{sgn}( k_z\zeta)}(\boldsymbol{k})\boldsymbol{\epsilon}_{ \mathrm{sgn}( k_z\zeta)}(\boldsymbol{k})e^{i\boldsymbol{k}\cdot\boldsymbol{x}}+\text{h.c}].\label{eq:A_T}
\end{align}
A corresponding mirror configuration is illustrated in Fig.~\ref{fig:mirrors}(b), where the two mirrors selectively reflect opposite helicity sectors of modes, reverse the helicity upon reflection, and are transparent to the complementary sector. The resulting PZW Hamiltonian is
\begin{align}
&H_{\rm P}^\zeta=
     \sum_\alpha\frac{[ \boldsymbol{p}_\alpha-q_\alpha  \boldsymbol{\hat{A}}_{\rm P}^\zeta(\boldsymbol{r}_\alpha)-q_\alpha \delta \boldsymbol{A}_{\rm P}^\zeta(\{\boldsymbol{r}_\alpha\})]^2}{2m_\alpha}+ \mathcal{E}^\text{Cou}\nonumber\\&+H^\text{rad}-\int d^3x \,\V{P}_\perp(\V{x})\cdot\V{\hat{E}}^\zeta(\V{x})+\sum_{\boldsymbol{k}}\frac{\mathcal{A}_k ^2\omega_k}{2\hbar}|\widetilde{\V{P}}_\perp(\boldsymbol{k})|^2,
\end{align}
where we define
\begin{align}
     \delta\boldsymbol{A}_{\rm P}^\zeta(\{\boldsymbol{r}_\alpha\})&=\zeta\sum_{\boldsymbol{k}}\frac{ \mathcal{A}_k ^2\text{sgn}(k_z)}{2\hbar}\bigg[\frac{\boldsymbol{k}}{|\boldsymbol{k}|}e^{i\boldsymbol{k}\cdot\boldsymbol{r}_\alpha}\nonumber\\&+i |\boldsymbol{k}|\int_0^1 u du \,e^{iu\boldsymbol{k}\cdot\boldsymbol{r}_\alpha} \boldsymbol{r}_\alpha\bigg]\times\widetilde{\V{P}}_\perp(\boldsymbol{k}).\label{eq:induce_A}
\end{align}
In the long-wavelength limit $\boldsymbol{k}\cdot\boldsymbol{r}_\alpha\to0$, the induced vector potential reduces to the dipole-associated form $\delta\boldsymbol{A}_{\rm P}^{\zeta}
\simeq
\zeta\sum_{\boldsymbol{k}}
\operatorname{sgn}(k_z)\mathcal{A}_k^2
\boldsymbol{k}\times\boldsymbol{d}/(2\hbar|\boldsymbol{k}|)$. For $N$ particles with the same charge $q$ and mass $m$, the leading-order Zeeman-like energy is 
 \begin{align}
    \Delta H^\zeta\simeq -\V{\mu}_R\cdot    \delta \boldsymbol{B}_{\rm P}^\zeta ,\label{eq:corr_T}
\end{align}
where $\V{\mu}_R=Nq \V{R}\times(-i\hbar\V{\nabla}_R)/(2m)$ is the collective orbital magnetic moment associated with the center-of-mass motion $\V{R}=\sum_\alpha\V{r}_\alpha/N$, and $\delta \boldsymbol{B}_{\rm P}^\zeta=q\zeta\sum_{\boldsymbol{k}}\text{sgn}(k_z)\mathcal{A}_k ^2 \boldsymbol{k}/(\hbar |\boldsymbol{k}|)$ is the induced Zeeman-like field. 

\textit{Two-body system with opposite charges in a two-mode spatial-chiral cavity}.—To understand the correction induced by a spatial-chiral vacuum, we consider a two-body model consisting of two oppositely charged particles confined in a spatial-chiral cavity. We denote their masses by $m_1$ and $m_2$, and their charges by $+q$ and $-q$, respectively, with $q>0$. For simplicity, we assume that the cavity supports only two degenerate helical modes with wavevectors $\V{k}$ and $-\V{k}$~\cite{Nonperturbative2023}. In addition, a harmonic trapping potential is applied to confine the center of mass with characteristic frequency $\omega_0$. In terms of the center-of-mass and relative coordinates, the PZW Hamiltonian takes the approximate form
\begin{align}
  &   h^\eta= \frac{ \boldsymbol{p}_{\V{R}}^{\,2}}{2M}+\frac{ \boldsymbol{p}_{\V{r}}^{\,2}}{2 m}+\mathcal{E}^\text{Cou}+\frac{M \omega_0^2 \V{R}^2}{2}+\frac{q^2(r^{x2}+r^{y2})}{2\varepsilon_0 V}  +\Delta h^\eta\nonumber\\
     &-\V{d}\cdot\hat{\boldsymbol{E}}^\eta(0)-\frac{\V{Q}\mathbin{:}\nabla\V{\hat{E}}^\eta(0)}{2}+\hbar\omega_k(a^\dagger_{\V{k}}a_{\V{k}}+a^\dagger_{-\V{k}}a_{-\V{k}}+1),\label{eq:h_eta}
\end{align}
where $\V{R}=(m_1\V{r}_1+m_2\V{r}_2)/M$ is the center of mass, $M=m_1+m_2$ is the total mass, $m=m_1m_2/M$ is the reduced mass and the dipole operator is given by $\V{d}=q\V{r}$. The inversion-breaking self-energy Eq.~(\ref{eq:corr_I}) becomes
\begin{align}
    \Delta h^\eta=\frac{q^2\eta k r^z}{2\varepsilon_0 V}(r^x R^y-r^y R^x).
\end{align}
For a hydrogen-like atomic $nS_{1/2}$ state, the leading nonvanishing energy shift induced by $\Delta h^\eta$ is second order and can be expressed in terms of the static scalar electric-quadrupole polarizability $\alpha_{E2}^{(0)}(nS_{1/2})$ in the low-trap-frequency limit~\cite{kharchenko2014,supplementary},
\begin{align}
    \Delta \mathcal{E}_{nS}^{(2)} = - \frac{1}{3}\left( \frac{q^2 k  \ell_R}{2\varepsilon_0 V} \right)^2
  \alpha_{E2}^{(0)}(nS_{1/2}),
    \label{eq:chiral_shift_polarizability}
\end{align}
where $\ell_R=\sqrt{\hbar/(2M\omega_0)}$. 

Alkali Rydberg states can exhibit sizable self-energy corrections because their extended wavefunctions and small level spacings give rise to large quadrupole polarizabilities~\cite{supplementary}. As a concrete example, we consider the Cs $6S_{1/2}\rightarrow8S_{1/2}$ two-photon transition. Using the alkali Rydberg Calculator package~\cite{Sibalic2017ARC}, we obtain $\alpha_{E2}^{(0)}(6S_{1/2})\simeq1.1\times10^4$ a.u. and $\alpha_{E2}^{(0)}(8S_{1/2})\simeq1.8\times10^7$ a.u. For a spatial-chiral cavity with $\lambda=100\,{\rm nm}$, $V=(\lambda/2)^3$, and $k\ell_R=0.1$, the resulting differential transition shift is $-22\,{\rm kHz}$. The same transition has been measured using two-photon frequency-comb spectroscopy with frequency sensitivity of the same order of magnitude~\cite{Fendel2007,Stalnaker2010femtosecond,kim2018direct}. This comparison indicates that the inversion-breaking self-energy can lead to experimentally measurable spectroscopic consequences and therefore should be included in the multipolar Hamiltonian for spatial-chiral photonic vacuums.

\textit{Zeeman correction in a temporal-chiral cavity}.—We next illustrate the consequence of the additional Zeeman-like field generated by the PZW transformation in a temporal-chiral cavity. We consider a two-dimensional harmonic oscillator confined in a plane and coupled to a single chiral cavity mode. To reduce the photonic degrees of freedom, we choose a single standing-wave mode with circular polarization. For a suitable choice of the confinement plane~\cite{supplementary}, the vector potential reduces to
\begin{align}
   \V{\hat{A}}^\zeta(x,y)=
   \mathcal{A}_0
   \left[
   \frac{\V{e}_x-i\zeta\V{e}_y}{\sqrt{2}}\,a_\zeta
   +{\rm h.c.}
   \right],
   \label{eq:A_t_model}
\end{align}
where $\zeta=\pm1$ denotes the cavity chirality and $\mathcal A_0=\sqrt{\hbar/(\varepsilon_0\omega_cV_c)}$. The PZW Hamiltonian in the dipole approximation becomes
\begin{align}
   h^\zeta&=\frac{[\V{p}-q\delta\V{A}^\zeta_{\rm P}(\boldsymbol{x})]^2}{2m}+\frac{m\omega^2\rho^2}{2}+\frac{q^2 \rho^{2}}{2\varepsilon_0 V}\nonumber\\
    &-q \V{x}\cdot\V{\hat{E}}^\zeta (0)+\hbar\omega_c(a_\zeta^\dagger a_\zeta+\frac{1}{2}),\label{eq:H_PZW_model}
\end{align}
with $\rho=\sqrt{x^2+y^2}$ and $ \delta\V{A}^\zeta_{\rm P}(\V{x}) =  \frac{q\zeta}{2\varepsilon_0\omega_cV}(x\V{e}_y-y\V{e}_x)$. Compared with the conventional symmetric PZW Hamiltonian, this induced vector potential gives the correction as follows
\begin{align}
    -\frac{q\delta\V{A}^\zeta_{\rm P}\cdot\V{p}}{m}
    +\frac{q^2(\delta\V{A}^\zeta_{\rm P})^2}{2m}
    =
    -\frac{\hbar\zeta l_z}{2}\frac{g^2}{\omega_c}
    +\frac{m g^4\rho^2}{8\omega_c^2},
    \label{eq:Zeeman_corr}
\end{align}
where $l_z=\V{e}_z\cdot(\V{x}\times\V{p})/\hbar$ and $g=|q|\mathcal{A}_0\sqrt{\omega_c/(m\hbar)}$ is the coupling strength. The first term is chirality- and angular-momentum dependent~\cite{jiang_angular2025}. 

\begin{figure}
    \centering
    \includegraphics[width=0.75\linewidth]{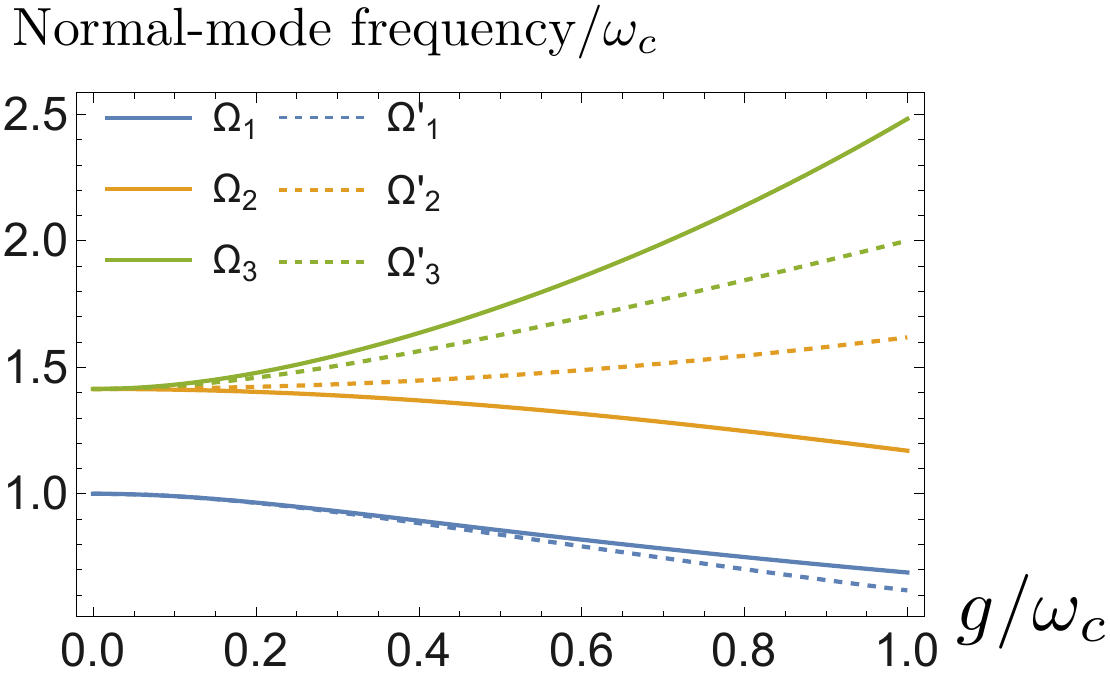}
    \caption{Normal-mode frequencies $\Omega_n$ and $\Omega_n'$ ($n=1,2,3$) for the full Hamiltonian in Eq.~(\ref{eq:H_PZW_model}) and the reference Hamiltonian without the dipole-associated Zeeman correction Eq.~(\ref{eq:Zeeman_corr}) obtained by setting $\delta\boldsymbol A_{\rm P}^{\zeta}=0$, respectively. The parameters are $\hbar=1$, $q=-e$, $\zeta=+1$, $\omega_c=1$, and $\omega=\sqrt{2}\omega_c$.}
    \label{fig:comparison}
\end{figure}

We diagonalize Eq.~\eqref{eq:H_PZW_model} by a Hopfield transformation and compare its normal-mode frequencies $\Omega_n$ with those of a reference Hamiltonian $h^{\prime\zeta}$ obtained by setting $\delta\V{A}^\zeta_{\rm P}=0$~\cite{hopfield,Rokaj2023,Liuchiral2024,supplementary}. Figure~\ref{fig:comparison} shows the resulting normal-mode frequencies as functions of the light--matter coupling strength $g$. As $g$ increases, the Zeeman correction becomes pronounced, especially for the $n=2,3$ modes in the strong-coupling regime $g/\omega_c\gtrsim0.3$~\cite{RMP2019,Ashida2021,Rubio2017,Rotation2023,Nonperturbative2023,Rubio2025,Rubio2025PRR,Rokaj2025}. The opposite signs of the frequency shifts originate from the angular-momentum-dependent term $-\hbar\zeta l_zg^2/(2\omega_c)$, demonstrating that the temporal-chiral PZW correction must be included for reliable spectral calculations in the strong-coupling regime.

As a concrete many-electron example, we consider a two-dimensional electron gas coupled to a single-mode chiral cavity. The resulting collective coupling strength between the center of mass coordinate and the cavity mode becomes $g=\sqrt{e^2n_e/(\varepsilon_0mL_z)}$, where $n_e$ is the electron area density and $L_z$ is the effective cavity length along the $z$ direction~\cite{Liuchiral2024}. For an electron density $n_e=2\times10^{15}\,{\rm m}^{-2}$~\cite{cavityQHE2022}, effective cavity length $L_z=80\,\mu{\rm m}$, effective mass $m=0.06m_e$ ($m_e$ is the bare electron mass), and cavity frequency $\omega_c=1\,{\rm THz}$, the resulting collective coupling satisfies $g/\omega_c\simeq1.15$. The corresponding relative frequency shifts in Fig.~\ref{fig:comparison} are $(\Omega_n-\Omega_n')/\Omega_n\simeq14\%,-52\%,24\%$ for $n=1,2,3$, respectively. These sizable shifts demonstrate that the additional temporal-chiral correction can become significant in realistic many-electron systems.

\textit{Conclusions}.—In this work, we applied the Power-Zienau-Woolley transformation to derive the multipolar light-matter Hamiltonian in symmetry-breaking photonic vacuums. We showed that, when the photonic vacuum breaks spatial inversion or time-reversal symmetry, the resulting Hamiltonian acquires additional terms beyond its conventional form in a symmetry-preserving vacuum.

For a spatial-chiral vacuum, the correction takes the form of an inversion-breaking self-energy. In a trapped hydrogen-like atomic system, this term gives rise to a second-order energy shift whose magnitude is comparable to the resolution of modern precision spectroscopy. For a temporal-chiral vacuum, the PZW transformation instead generates a cavity-induced Zeeman-like field. Through a charged two-dimensional harmonic oscillator model, we see that coupling to a single temporal-chiral cavity mode leads to an angular-momentum-dependent spectral correction.


Taken together, our results demonstrate that the symmetry of the surrounding photonic vacuum qualitatively modifies the multipolar Hamiltonian, providing a framework for describing light-matter coupling in chiral cavity QED. In the future, it would also be valuable to extend general gauge formulations of light-matter Hamiltonians to photonic vacuums with broken inversion or time-reversal symmetry~\cite{Unifying1987,Gauge1990,kumar2010,Stokes2012,Stokes_2018,Chou_2019,stokes2019gauge,Stokes2020,StokesRMP,Stokes2023,Huo2020,Huo2025}.

\textit{End Matter}.—
\begin{table*}[t] 
\centering 
\begin{tabular}{|c|c|c|} \hline \textbf{Property} & \textbf{Spatial-Chiral Cavity}& \textbf{Temporal-Chiral Cavities}\\ \hline Vector potential& $\boldsymbol{\hat{A}}^{\eta}(\boldsymbol{x})$ (Eq.~(\ref{eq:A_I}))&$\boldsymbol{\hat{A}}^{\zeta}(\boldsymbol{x})$ (Eq.~(\ref{eq:A_T}))\\ \hline Polarization& $ \boldsymbol{\epsilon}_\eta(\boldsymbol{k} )= \boldsymbol{\epsilon}_\eta^\ast(-\boldsymbol{k})$& $\bar{\boldsymbol{\epsilon}}_\zeta(\boldsymbol{k})=\bar{\boldsymbol{\epsilon}}_\zeta(-\boldsymbol{k})$\\ \hline Asymmetry& $\hat{\mathcal{I}}\boldsymbol{\hat{A}}^{\eta}(\boldsymbol{x})\hat{\mathcal{I}}^{-1}=-\boldsymbol{\hat{A}}^{-\eta}(-\boldsymbol{x})\neq-\boldsymbol{\hat{A}}^{\eta}(-\boldsymbol{x})$& $\hat{\mathcal{T}}\boldsymbol{\hat{A}}^{\zeta}(\boldsymbol{x})\hat{\mathcal{T}}^{-1}=-\boldsymbol{\hat{A}}^{-\zeta} (\boldsymbol{x})\neq-\boldsymbol{\hat{A}}^{\zeta}(\boldsymbol{x})$\\ \hline Induced field& Electric-like $ \delta\boldsymbol{E}^{\eta}_{\rm P}(\{\boldsymbol{r}_\alpha\})$ (Eq.~(\ref{eq:deltaE}))& Zeeman-like $\delta \boldsymbol{B}_{\rm P}^\zeta$ (Eq.~(\ref{eq:corr_T}))\\ \hline Lowest-order correction& $-\boldsymbol{d}\cdot\delta\boldsymbol{E}^{\eta}_{\rm P}$ (Eq.~(\ref{eq:corr_I}))& $-\V{\mu}_R\cdot \delta \boldsymbol{B}_{\rm P}^\zeta $ (Eq.~(\ref{eq:corr_T}))\\\hline 
\end{tabular} 
\caption{Comparison of the properties of spatial and temporal chiral cavities.} \label{table:comparison} 
\end{table*}
By analogy with conventional dipole coupling to an external electric field, the quantity
\begin{align} \delta\boldsymbol{E}^{\eta}_{\rm P} = \sum_{\boldsymbol{k}} \frac{\eta \mathcal{A}_k^2\omega_k}{2\hbar|\boldsymbol{k}|} \left(\boldsymbol{k}\times\boldsymbol{Q}\cdot\boldsymbol{k}\right)\label{eq:deltaE} \end{align}
in Eq.~(\ref{eq:corr_I}) can be interpreted as an effective cavity-induced electric field generated by the quadrupole moment. In contrast, the temporal-chiral vacuum induces a multipole-dependent vector potential, which reduces in the dipole approximation to the Zeeman-like correction in Eq.~\eqref{eq:corr_T}. The comparison between the two cases is summarized in Table~\ref{table:comparison}.

\textit{Acknowledgments}.—
We acknowledge helpful discussions with Ahsan Nazir, Yiming Pan, Gabriel Cardoso, Junxiao Hui and Thors Hans Hansson. This work was supported by National Natural Science Foundation of China (NSFC) under Grant No. 12374332,  Jiaoda 2030 program WH510363001-1, the Innovation Program for Quantum Science and Technology Grant No. 2021ZD0301900.

\bibliography{ref.bib}

\end{document}